\documentclass{aa}
\usepackage{graphicx}
\usepackage{amssymb}

\usepackage{txfonts}

\newcommand{\mmean}{$\langle \dot{M} \rangle$}
\newcommand{\mmax}{$\dot{M}_\mathrm{max}$}
\newcommand{\mach}{${\mathcal M}$}
\newcommand{\etal}{et al.}

\begin{document}

\title{Protostellar mass accretion rates\\from gravoturbulent fragmentation}

   \author{S. Schmeja
          \and
          R. S. Klessen
          }

   \offprints{S. Schmeja}

   \institute{Astrophysikalisches Institut Potsdam,
              An der Sternwarte 16, D-14482 Potsdam, Germany\\
              \email{sschmeja@aip.de, rklessen@aip.de}}

   \date{Received ... / accepted ...}

   \abstract{ We analyse protostellar mass accretion rates $\dot{M}$
     from numerical models of star formation based on gravoturbulent
     fragmentation, considering a large number of different environments.
     To within one order of magnitude, $\dot{M} \approx M_\mathrm{J} / \tau_\mathrm{ff}$
     with $M_\mathrm{J}$ being the mean thermal Jeans mass and $\tau_\mathrm{ff}$
     the corresponding free-fall time.
     However,
     mass accretion rates are highly time-variant,
     with a sharp peak shortly after the formation of the protostellar
     core.
     We present an empirical exponential fit formula to describe the time evolution
     of the mass accretion and discuss the resulting fit parameters.
     There is a positive correlation between the peak accretion rate
     and the final mass of the protostar. We also investigate the
     relation of $\dot{M}$ with the turbulent flow velocity as well as
     with the driving wavenumbers in different environments.  We then
     compare our results with other theoretical models of star
     formation and with observational data.
   }

   \maketitle
%

\section{Introduction}
\label{sec:intro}

Stars are born in dense cores of interstellar molecular clouds.
Despite recent observational and theoretical progress, the initial
conditions and physical processes that determine the formation of
stars are still not fully understood.

In the so-called ``standard theory of star formation'' (Shu \etal\
\cite{sal87}) stars are formed by the inside-out collapse of a
singular isothermal sphere that is initially in quasistatic
equilibrium, supported against gravity by magnetic and thermal
pressure and evolves only due to slow ambipolar diffusion processes.
This model predicts protostellar mass accretion rates that are
constant with time and only depend on the isothermal sound speed (Shu
\cite{shu77}).  This hypothesis, however, has been challenged from
several sides (see Larson \cite{larson03} or Mac~Low \& Klessen 
\cite{mor_ralf} for a summary).
It only is applicable to isolated, single stars, while it is known
that the majority of stars form in small aggregates or large clusters
(Adams \& Myers \cite{adams_myers}; Lada \& Lada \cite{ladalada}).
Furthermore, there is both observational evidence (Crutcher
\cite{crutcher}; Andr{\'e} \etal\ \cite{andre00}; Bourke \etal\
\cite{bourke}) and theoretical reasoning (e.g.\ Nakano
\cite{nakano98}) showing that most observed cloud cores do not have
magnetic fields strong enough to support against gravitational
collapse.  Similarly, the long lifetimes implied by the quasi-static
phase of evolution in the model are difficult to reconcile, e.g., with
observational statistics of cloud cores (Taylor \etal\ \cite{tmw96};
Lee \& Myers \cite{lee_my}; Visser \etal\ \cite{vrc02}) and with
chemical age considerations (van Dishoeck \& Blake \cite{dis_blake};
Langer \etal\ \cite{langer00}).

  Molecular clouds appear to actually be transient objects with
  lifetimes of a few million years that form and dissolve in the
  larger scale turbulent flow of the Galactic disc
  (Ballesteros-Paredes \etal\ \cite{bhv99}; Elmegreen \cite{elmeg00};
  Hartmann \etal\ \cite{hbb01}; Hartmann \cite{hartmann03};
  V\'azquez-Semadeni \etal\ \cite{vsb03}).  Observations of
  self-similar structure in molecular clouds (e.g. Mac Low \&
  Ossenkopf \cite{maclow_oss00}; Ossenkopf \& Mac Low
  \cite{oss_maclow02}) indicate that interstellar turbulence is driven
  on scales substantially larger than the clouds themselves.  These
  large-scale turbulent flows compress and cool gas. At sufficiently
  high densities atomic gas is then quickly converted into molecular
  form (Hollenbach \etal\ \cite{hws71}).  These same
  flows will continue to drive the turbulent motions observed within
  the newly formed cloud.  Some combination of turbulent flow, free
  expansion at the sound speed of the cloud, and dissociating
  radiation from internal star formation will then be responsible for
  their destruction on a timescale of 5--10~Myr.  The most likely
  source of such large-scale interstellar turbulence in the Milky Way
  is the combined energy and momentum input from supernovae
  explosions. They appear to overwhelm all other possibilities.  In
  the outer reaches of the Galaxy and in low surface brightness
  galaxies, on the other hand, the situation is not so clear, with
  magnetorotational or gravitational instabilities looking most likely
  to drive the observed flows (Mac~Low \cite{maclow02}; Mac~Low \&
  Klessen \cite{mor_ralf}).

Modern star formation theory, therefore, considers supersonic
interstellar turbulence as controlling agent for stellar birth, rather
than mediation by magnetic fields (Mac~Low \& Klessen
\cite{mor_ralf}).
This turbulence typically carries sufficient energy to counterbalance
gravity on global scales. On small scales, however, it may actually
provoke localised collapse (Hunter \& Fleck \cite{hunter_fleck}; Elmegreen \cite{elmeg93};
Padoan \cite{padoan95}; Ballesteros-Paredes \etal\ \cite{bvs99};
Klessen \etal\ \cite{khm00}; Padoan \& Nordlund \cite{pado_nord99},
\cite{pado_nord02}).  This apparent paradox can be resolved when
considering that supersonic turbulence establishes a complex network
of interacting shocks, where converging flows generate regions of
enhanced density.
%
%
The system becomes highly filamentary, with elongated structures
  being caused either by shear motions or by compression at the
  intersection of shocked layers of gas. At some locations the density
  enhancement can be sufficiently strong for gravitational instability
  to set in. The stability criteria for filaments and sheets have been
  derived and discussed in the context of star formation, e.g., by
  Larson (\cite{larson85}), Lubow \& Pringle (\cite{lp93}), and Clarke
  (\cite{clarke99}).  However, the same random flow that creates
  density enhancements may disperse them again.  For local collapse to
  result in stellar birth, it must progress fast enough for the region
  to `decouple' from the flow.

The efficiency of protostellar core
formation, the growth rates and final masses of the protostars, and
essentially all other properties of nascent star clusters then depend
on the intricate interplay between gravity on the one hand side and
the turbulent velocity field in the cloud on the other. The star
formation rate is regulated not just at the scale of individual
star-forming cores through ambipolar diffusion balancing magnetostatic
support, but rather at all scales (Elmegreen \cite{elmeg02}), via the
dynamical processes that determine whether regions of gas become
unstable to prompt gravitational collapse. The presence of magnetic
fields does not alter that picture significantly (Mac~Low \etal\
\cite{maclow98}; Stone \etal\ \cite{sog98}; Padoan \& Nordlund
\cite{pado_nord99}; Heitsch \etal\ \cite{heitsch01}). In particular,
it cannot prevent the decay of interstellar turbulence.

Clusters of stars build up in molecular cloud regions where
self-gravity overwhelms turbulence, either because such regions are
compressed by a large-scale shock, or because interstellar turbulence
is not replenished and decays on short timescales.  Then, many gas
clumps become gravitationally unstable synchronously to go into
collapse. If the number density is high, contracting protostellar
cores interact and may merge to produce new cores which now contain
multiple protostars.  Close encounters drastically alter the
trajectories of the protostars, thus changing their mass accretion
rates. This has important consequences for the final stellar mass
spectrum (Bonnell \etal\ \cite{bonnell97}; Klessen \& Burkert \cite{kless_bur00},
\cite{kless_bur01}; Bonnell \etal\ \cite{bonnell01a}, \cite{bonnell01b};
Klessen \cite{klessen01b}; Bate \etal\ \cite{bbb02}).

Inefficient, isolated star formation will occur in regions which are
supported by turbulence carrying most of its energy on very small
scales. This requires an unrealistically large number of driving
sources and appears at odds with the measured velocity structure in
molecular clouds which in almost all cases is dominated by large-scale
modes (Mac Low \& Ossenkopf \cite{maclow_oss00}; Ossenkopf \& Mac Low
\cite{oss_maclow02}).

In this paper we extend the analysis of protostellar mass accretion rates
from a single case (Klessen \cite{klessen01a}) to a large
series of numerical models of turbulent molecular cloud fragmentation,
which essentially cover the entire spectrum of observed star-forming
regions, ranging from inefficient and isolated star formation to the
fast and efficient build-up of stellar clusters.  These calculations,
their numerical realisation, and the adopted parameters are described
in Section \ref{sec:models}.  In Section \ref{sec:discussion} we
discuss our findings. We investigate the mass growth history of all
protostars in our set of models and present a simple analytic fit formula
for the accretion rate $\dot{M}$. We discuss our study in relation
with previous analyses and observational data
 in Sections \ref{sect:comp_mod} and \ref{sect:obs}, respectively.
 Finally, in Section \ref{sec:summary} we summarise our results.


\section {The models}
\label{sec:models}

To adequately describe the fragmentation of turbulent,
self-gravitating gas clouds, and the resulting formation and mass
growth of protostars, it is prerequisite to resolve the dynamical
evolution of collapsing cores over several orders of magnitude in
density. Due to the stochastic nature of supersonic turbulence, it is
not known in advance where and when this local collapse occurs. Hence,
SPH ({\em smoothed particle hydrodynamics}) is used to solve the
equations of hydrodynamics. It is a Lagrangian method, where the fluid
is represented by an ensemble of particles and flow quantities are
obtained by averaging over an appropriate subset of the SPH particles
(Benz \cite{benz90}; Monaghan \cite{monagh92}). The method is able to
resolve large density contrasts as particles are free to move and so
naturally the particle concentration increases in high-density
regions.
%

We use the same smoothing procedure for gravity and pressure
forces. This is one requirement to prevent artificial fragmentation
(Bate \& Burkert \cite{bb97}).
 Because it is computationally
prohibitive to treat the cloud as a whole, we concentrate on
subregions within the cloud and adopt periodic boundary conditions
(Klessen \cite{klessen97}). Once the central region of a collapsing
protostellar core exceeds a density contrast of $\sim 10^5$, it is
replaced by a ``sink'' particle (Bate \etal\ \cite{bbp95}), which has
the ability to accrete gas from its surrounding while at the same time
keeping track of mass and linear and angular momentum. By adequately
replacing high-density cores with sink particles we can follow the
dynamical evolution of the system over many free-fall times.

\begin{table*}[t]
\caption{Overview of our models (See text for details)}
    \label{tab:models}
{\small
\begin{center}
\begin{minipage}{15cm}
\begin{tabular}{l c c r l r r r r  r r r r}
\hline
\hline
Name
& \mach 
& $k$ 
& \multicolumn{1}{c}{$n_\mathrm{p}$\footnote{number of particles in the
simulation}}
& $M_\mathrm{min}$\footnote{SPH resolution limit}
& $M_\mathrm{accr}$\footnote{fraction of the total mass that has been accreted by the end of the simulation}
& $n_*$\footnote{total number of formed protostars}
& $n_*$\footnote{number of protostars that can be fitted by Eq.~(\ref{eq:fit})} & $\sigma_\mathrm{mean}$\footnote{mean deviation of the fits, calculated from Eq.~(\ref{eq:sigma})}
& \multicolumn{4}{c}{$\dot{M}_\mathrm{mean}$ [$10^5 \mathrm{M_{\sun} yr^{-1}}$]}\\
& & & & \multicolumn{1}{c}{\scriptsize [M$_{\sun}$]} & \multicolumn{1}{c}{\scriptsize [\%]} &  & {\scriptsize fitted} & & {\scriptsize bin1} & {\scriptsize bin2} & {\scriptsize bin3} & {\scriptsize bin4} \\
\hline
G1    & --& --&  50\,000 & 0.44  & 93.1 & 56 & 31 & 0.49 & 1.18 & 1.30 & 1.70 & 2.93 \\
G2    & --& --& 500\,000 & 0.044 & 84.9 & 56 & 52 & 0.43 & 0.94 & 1.40 & 2.09 & 3.51\vspace{0.2cm}\\
M01k2 & 0.1 & 1..2 & 205\,379 & 0.058 & 74.9 & 95 & 91 & 0.43 & 0.77 & 1.76 & 3.04 & --    \\ 
M01k4 & 0.1 & 3..4 & 205\,379 & 0.058 & 27.2 &  3 &  3 & 0.81 & 2.83 & --   & --   & 57.98 \\ 
M01k8 & 0.1 & 7..8 & 205\,379 & 0.058 & 85.9 &  3 &  3 & 0.47 & --   & 1.37 & 13.25& 59.16 \\ 
M05k2 & 0.5 & 1..2 & 205\,379 & 0.058 & 37.2 & 23 & 22 & 0.49 & 1.63 & 5.05 & 4.88 & 12.95 \\ 
M05k4 & 0.5 & 3..4 & 205\,379 & 0.058 & 77.9 & 48 & 48 & 0.39 & 1.31 & 2.49 & 2.56 & 6.30  \\ 
M05k8 & 0.5 & 7..8 & 205\,379 & 0.058 & 59.5 & 48 & 48 & 0.40 & 1.34 & 2.22 & 3.79 & 7.77  \\ 
M2k2  &  2  & 1..2 & 205\,379 & 0.058 & 75.1 & 68 & 62 & 0.41 & 0.86 & 1.38 & 2.54 & 4.32  \\ 
M2k4  &  2  & 3..4 & 205\,379 & 0.058 & 47.9 & 62 & 62 & 0.44 & 1.35 & 1.92 & 2.43 & 3.84  \\ 
M2k8  &  2  & 7..8 & 205\,379 & 0.058 & 66.2 & 42 & 40 & 0.42 & 0.87 & 1.29 & 1.72 & 3.38  \\ 
M3k2  & 3.2 & 1..2 & 205\,379 & 0.058 & 79.7 & 65 & 65 & 0.46 & 1.31 & 1.86 & 2.98 & 3.78  \\ 
M3k4  & 3.2 & 3..4 & 205\,379 & 0.058 & 82.1 & 37 & 35 & 0.55 & 1.01 & 1.13 & 1.15 & 1.86  \\ 
M3k8  & 3.2 & 7..8 & 205\,379 & 0.058 & 60.2 & 17 & 17 & 0.41 & 0.51 & 1.09 & 1.74 & 5.84  \\ 
M6k2a &  6  & 1..2 & 205\,379 & 0.058 & 85.4 &100 & 97 & 0.49 & 0.79 & 1.69 & 1.96 & 3.39  \\ 
M6k4a &  6  & 3..4 & 205\,379 & 0.058 & 62.4 & 98 & 93 & 0.44 & 0.38 & 0.83 & 0.89 & 1.10  \\ 
M6k2b &  6  & 1..2 & 195\,112 & 0.062 & 34.5 & 50 & 50 & 0.42 & 1.02 & 1.81 & 2.50 & --    \\ 
M6k4b &  6  & 3..4 &  50\,653 & 0.24  & 29.7 & 50 & 47 & 0.43 & 0.72 & 1.66 & 1.87 & --    \\ 
M6k8b &  6  & 7..8 &  50\,653 & 0.24  & 35.7 & 25 & 25 & 0.44 & 0.35 & 0.61 & 1.38 & 2.47  \\ 
M6k2c &  6  & 1..2 & 205\,379 & 0.058 & 75.8 &110 & 97 & 0.43 & 0.83 & 1.32 & 1.50 & 1.23  \\ 
M6k4c &  6  & 3..4 & 205\,379 & 0.058 & 61.9 & 53 & 46 & 0.54 & 1.31 & 1.46 & 2.05 & 1.97  \\ 
M6k8c &  6  & 7..8 & 205\,379 & 0.058 &  6.4 & 12 & 10 & 0.43 & 0.50 & 0.62 & 0.58 & --    \\ 
M10k2 & 10  & 1..2 & 205\,379 & 0.058 & 56.5 &150 &146 & 0.44 & 1.08 & 2.62 & 2.09 & --    \\ 
M10k8 & 10  & 7..8 & 205\,379 & 0.058 & 32.4 & 54 & 44 & 0.53 & 0.26 & 0.64 & --   & --    \\ 
\end{tabular}
\end{minipage}
\end{center}
}
\end{table*}

The suite of models consists of two globally unstable models that
contract from Gaussian initial conditions without turbulence (for
details see Klessen \& Burkert \cite{kless_bur00}, \cite{kless_bur01})
and of 22 models where turbulence is maintained with constant rms Mach
numbers $\cal M$, in the range $0.1 \le {\cal M} \le 10$. We
distinguish between turbulence that carries its energy mostly on large
scales, at wavenumbers $1 \le k \le 2$, on intermediate scales, i.e.\
$3 \le k \le 4$, and on small scales with $7 \le k \le 8$. The
corresponding wavelengths are $\ell = L/k$, where $L$ is the total
size of the computed volume.  The models are labelled mnemonically as
M$\cal M$k$k$, with rms Mach number $\cal M$ and wavenumber $k$, while
G1 and G2 denote the two Gaussian runs.  The main parameters are
summarised in Table~\ref{tab:models}.
%
%
To have well defined
environmental conditions given by $\cal M$ and $k$, $\cal M$ is
required to be constant throughout the evolution. However, turbulent
energy dissipates rapidly, roughly on a free-fall timescale (Mac Low
\etal\ \cite{maclow98}; Stone \etal\ \cite{sog98}; Padoan \& Nordlund
\cite{pado_nord99}). We therefore apply a non-local driving scheme
that inserts energy at a given rate and at a given scale $k$. We use
Gaussian random fluctuations in velocity. This is
appealing because Gaussian fields are fully determined by their power
distribution in Fourier space.  We define a cartesian mesh with $64^3$
cells, and for each three-dimensional wave number $\vec{k}$ we
randomly select an amplitude from a Gaussian distribution around unity
and a phase between zero and $2\pi$.  We then transform the resulting
field back into real space to get a ``kick-velocity'' in each cell.  Its
amplitude is determined by solving a quadratic equation such to keep
$\cal M$ constant (Mac~Low \cite{maclow99}; Klessen \etal\
\cite{khm00}). The ``kick-velocity'' is then simply added to the speed
of each SPH particle located in the cell. We adopted this method for
mathematical simplicity. In reality, the situation is far more
complex. Still, our models of large-scale driven clouds contain many
features of molecular clouds in supernovae driven turbulence (e.g.
Ballesteros-Paredes \& Mac~Low \cite{bpml02}; Mac~Low \etal\ \cite{mak03}).
Conversely, our models of small-scale turbulence
bear certain resemblance to energy input on small scales provided by
protostellar feedback via outflows and winds.

Our models neglect the influence of magnetic fields, because their
presence cannot halt the decay of turbulence (Mac~Low \etal\ 
\cite{maclow98}; Stone \etal\ \cite{sog98}; Padoan \& Nordlund
\cite{pado_nord99}) and does not significantly alter the efficiency of
local collapse for driven turbulence (Heitsch \etal\ \cite{hmk01}).
More importantly, we do not self-consistently consider feedback
effects from the star formation process itself (like bipolar outflows,
stellar winds, or ionising radiation from new-born O or B stars). Our
analysis of protostellar mass accretion rates solely focuses on the
interplay between turbulence and self-gravity only. This is also the
case in the Shu (\cite{shu77}) theory of isothermal collapse. Hence,
our findings can be directly compared to the ``standard theory of star
formation''.

The models are computed in normalised units using an isothermal
equation of state. Scaled to physical units we adopt a temperature of
11.3$\,$K corresponding to a sound speed $c_{\rm s} =
0.2\,$km$\,$s$^{-1}$, and we use a mean density of $n({\rm H}_2) =
10^5\,$cm$^{-3}$, which is typical for star-forming molecular cloud
regions (e.g.\ in $\rho$~Ophiuchi, see Motte \etal\ \cite{man98}).
The total mass contained in the computed volume in the two
Gaussian models is 220$\,$M$_{\sun}$ and the size of the cube is
$0.34\,$pc. This corresponds to 220 thermal Jeans masses. The
turbulent models have a mass of 120$\,$M$_{\sun}$ within a volume of
($0.28\,{\rm pc})^3$, equivalent to 120 thermal Jeans
masses\footnote{We use a spherical definition of the
  Jeans mass, $M_{\rm J} \equiv 4/3\,\pi \rho (\lambda_{\rm
  J}/2)^3$, with density $\rho$ and Jeans length $\lambda_{\rm J}\equiv
  \left(\frac{\pi{\cal R}T }{G \rho}\right)^{1/2}$ and where $G$ and
  $\cal R$ are the gravitational and the gas constant. The mean Jeans
  mass $\langle M_{\rm J} \rangle$ is then determined from the average
  density in the system $\langle \rho \rangle$.
  }. The mean thermal Jeans mass in all models is thus $\langle M_{\rm J}
\rangle = 1\,$M$_{\sun}$, the global free-fall timescale is $\bar{\tau}_{\rm ff} =
10^5\,$yr, and the simulations cover a density range from $n({\rm H}_2)
\approx 100\,$cm$^{-3}$ in the lowest density regions to $n({\rm H}_2) \approx
10^9\,$cm$^{-3}$ where collapsing protostellar cores are identified and
converted into ``sink'' particles in the code.  This coincides in time with
the formation of the central protostar to within $\sim10^3\,$yr (Wuchterl \&
Klessen \cite{wucht_kless01}).
%
The resolution limit for each model, requiring that the local Jeans mass
is always resolved by at least 100 gas particles (Bate \& Burkert \cite{bb97}),
is given in Col.~5 of Table~\ref{tab:models}.

%
In the subsequent protostellar phase of evolution, we determine
  accretion rates $\dot{M}$ by measuring the amount of mass as
  function of time that falls into a control volume defined by each
  ``sink'' particle.  Its diameter is fixed to $560\,$AU. Entering gas
  particles pass through several tests to check if they remain bound
  to the ``sink'' particle (Bate \etal\ \cite{bbp95}) before they are
  considered accreted. As all gas particles have the same  mass and as
  accretion events occur at random times, the resulting accretion
  rates are mass-binned and we smooth over a few consecutive accretion
  events to get a description of the time evolution of $\dot{M}$.
We cannot resolve the evolution in the interior of the control volume.
%
  Because of angular momentum conservation most of the matter that falls in
  will assemble in a protostellar disc. There it is transported inwards by
  viscous and possibly gravitational torques. The latter will be provided by
  spiral density waves that develop when the disc becomes too massive, which
  happens when mass is loaded onto the disc faster than it is removed by
  viscous transport alone. Altogether, the disc will not prevent or delay material
  from accreting onto the protostar for long. It acts as a buffer and smoothes
  eventual accretion spikes.
For the mass range considered here also feedback effects are too
weak to halt or delay accretion.  With typical disc sizes of order of several
hundred AU, the control volume therefore fully encloses both, star and disc,
and the measured core accretion rates are good estimates for the actual
stellar accretion rates.  Deviations may be expected only if the protostellar
core forms a binary star, where the infalling mass must then be distributed
between two stars, or if very high-angular momentum material is accreted,
where a certain mass fraction may end up in a circumbinary disc and not
accrete onto a star at all.

In the prestellar phase, i.e.\ before the central
protostar forms, we determine the accretion history by computing the change
of mass inside the control volume centered on the SPH particle that
turns into a ``sink'' during the later evolution. Turbulent compression
leads to mass growth, i.e.\ $\dot{M}>0$, while expansion will result
in mass loss and $\dot{M}<0$. Appreciable mass growth, however, is
only achieved when gravity takes over and the region goes into
collapse.




\section{Discussion}
\label{sec:discussion}

\subsection{First approximation to $\dot{M}$}

The entire process of molecular cloud collapse and build-up of the stellar
cluster lasts several global free-fall times ($\bar{\tau}_\mathrm{ff} = 10^5$~yr).
Likewise, the accretion process of a protostellar core takes place on a timescale
of a few $\bar{\tau}_\mathrm{ff}$, comparable to most other models of star formation.

A simple approximation to the accretion rate can be achieved by dividing the local
Jeans mass by the local dynamical timescale:
\begin{equation}
\dot{M} \approx M_\mathrm{J}/\tau_\mathrm{ff}
\label{eq:jeans}
\end{equation}
By substituting
\begin{equation}
M_\mathrm{J} = \frac{\pi^{5/2}}{6} \rho_0^{-1/2} \left(\frac{\mathcal{R} T}{G}\right)^{3/2}
 =  \frac{\pi^{5/2}}{6} \rho_0^{-1/2} G^{-3/2} c_\mathrm{s}^3,
 \end{equation}
where $\rho_0$ denotes the initial density, $T$ the temperature and $c_\mathrm{s}$
the iso\-thermal sound speed, and
 \begin{equation}
 \tau_\mathrm{ff} = \sqrt{\frac{3\,\pi}{32\,G \rho_0}}
 \end{equation}
Eq.~(\ref{eq:jeans}) can be written as
\begin{equation}
\dot{M} \approx \frac{M_\mathrm{J}}{\tau_\mathrm{ff}} =
\sqrt{\frac{32}{3}} \frac{\pi^2}{6} \frac{c_\mathrm{s}^3}{G} =
5.4\,\frac{c_\mathrm{s}^3}{G},
\end{equation}
depending only on the isothermal sound speed (or temperature).
For a sound speed $c_\mathrm{s} = 0.2~\mathrm{km\,s^{-1}}$ we obtain
$\dot{M} = 10^{-5}~\mathrm{M_{\sun}\,yr^{-1}}$.
This is higher than the accretion rate for the collapse of a singular isothermal
sphere: $\dot{M} = 0.975\,c_\mathrm{s}^3/G$ (Shu \cite{shu77}).
However, the accretion rates in our models vary with time.
Typical peak accretion rates are roughly in the range $(3-50)\,c_\mathrm{s}^3/G$
or $5 \times 10^{-6}$ to $10^{-4}~\mathrm{M_{\sun}\,yr^{-1}}$.
The values exceed the approximated value $M_\mathrm{J}/\tau_\mathrm{ff}$
due to external compression in the turbulent flow.

\begin{figure*}[t]
  \centering \includegraphics[width=17cm]{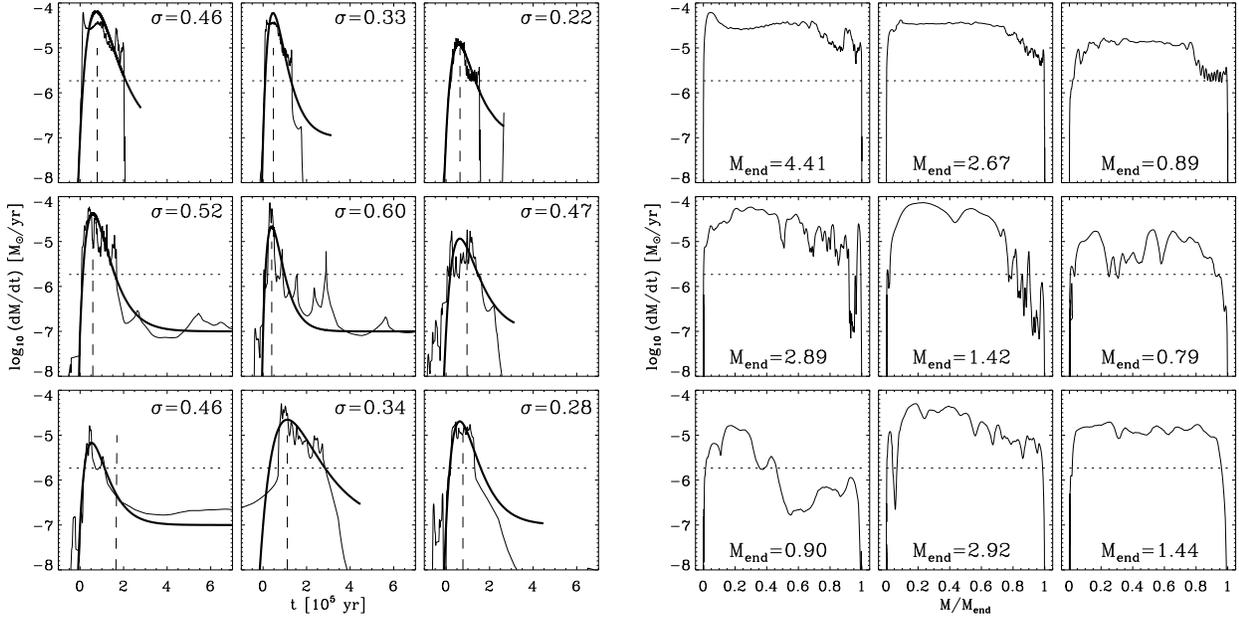}
   \caption{Mass accretion rates of nine randomly selected protostellar cores
     of three different models.
     Left panel: $\dot{M}$ versus time for a Gaussian collapse
     (G2; upper row), a turbulent model driven on a large scale (M6k2a;
     middle row), and a turbulent model driven on a small scale (M6k8b; lower row).
     The thin line represents the actual simulation, the thick line the fit
     as described in the text.
     The deviation $\sigma$ as given by Eq.~(\ref{eq:sigma}) is indicated
     for each object.
     The dotted line shows the constant accretion rate that would be expected from
     the classical Shu (\cite{shu77}) scenario.
     The dashed line stands for the assumed transition from Class~0 to Class~I.
     The right panel shows the same protostellar cores as on the left side plotted
     versus the ratio of accreted to final mass.
     The final masses (in M$_{\sun}$) are also given.}
   \label{accrcurve}
\end{figure*}

\subsection{Time-varying mass accretion rates}

We analyse the full mass growth history of all protostellar cores in
our models and
 we find that mass accretion rates from gravoturbulent
fragmentation are highly time-variable.
Several examples of the accretion rate $\dot{M}$ are
displayed in Fig.~\ref{accrcurve}, plotted versus time
(left panel) and the ratio of accreted to final mass
(right panel), respectively.
The maximum accretion rate is reached
rather rapidly 
and is then followed by
a somewhat slower decline. In some cases this decline is interrupted by
one or more secondary peaks.
As shown above,
the maximum accretion rate is significantly higher than the constant rate
predicted by the classical isothermal collapse model (
plotted as dotted line in Fig.~\ref{accrcurve}), but it falls below
that value in later stages.
Due to the dynamical interaction and competition between protostellar cores,
the mass accretion rates of cores in a dense cluster are different from
those of isolated cores.
In the first stage a core accretes local gas from its immediate vicinity.
Once the local reservoir is depleted, the core may accrete fresh gas streaming
in from farther away or by encounters with non-collapsed gas clumps
(see discussion in Klessen \& Burkert \cite{kless_bur00}).
This results in secondary accretion peaks that are
also visible in the right panel of Fig.~\ref{accrcurve}, where one would expect
a single bump in the case of an isolated core.
For example, the central graph of the right panel of Fig.~\ref{accrcurve} nicely shows that
this particular protostar accretes only about half of the final mass
from its direct environment (first bump), while the rest stems from later accretion
events.

The transition phase between Class~0 and
Class~I protostars is believed to take place when about half of the
final mass has been accumulated (Andr\'{e} et al.\ \cite{andre00}).
This time is indicated by the dashed line in Fig.~\ref{accrcurve}.
Typically it takes place during or at the end of the peak accretion phase.
It determines the lifetime of Class~0 objects, which will be discussed below.

We define a mean accretion rate \mmean\ by averaging $\dot{M}$ in the
mass range $0.1 \le M/M_\mathrm{end} \le 0.8$, with $M_\mathrm{end}$
being the final mass of the protostar.  This phase
typically lasts only a few $10^4$ years. This is short compared to the
full accretion history. The bulk of stellar material is therefore
accumulated in the short time interval while the system is close to
maximum accretion.

\begin{figure*}
   \centering
   \includegraphics[width=17cm]{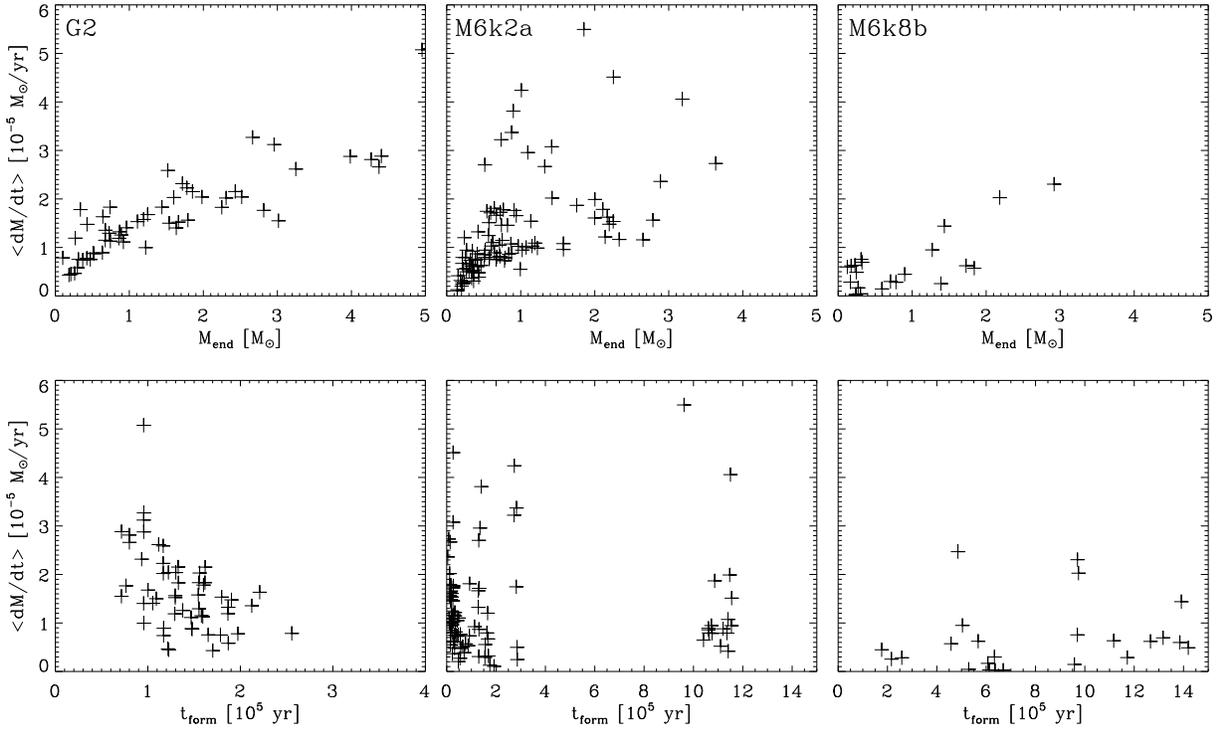}
   \caption{Mean accretion rates \mmean\ versus final mass (upper panel) and versus
   time of core formation (lower panel) for the same models as in Fig.~\ref{accrcurve}.
   The zero point of the timescale coresponds to the time when gravity is ``switched on''.
   Note the different timescales on which the formation of the cluster takes place.}

   \label{meanaccr}
\end{figure*}

In Fig.~\ref{meanaccr}, we plot the mean accretion rates versus final
star mass $M_\mathrm{end}$ and versus time of core formation
$t_\mathrm{form}$, respectively, for the same models as in
Fig.~\ref{accrcurve}.  Not surprisingly, \mmean\ increases with
increasing stellar mass, and decreases when the core forms later, although
this second correlation is not that clear.  In other words, more massive
stars have higher mass accretion rates and start to form first.
They can grow large, because on average they form in the high-density
regions of the cluster centre where they are able to maintain relatively
high accretion rates over a long time as more and more gas falls in from the
cluster outskirts.

\subsection{An empirical fit formula for $\dot{M}$}

One of our aims is to find a simple-to-use fit formula to approximate
the accretion process.  The protostellar mass growth history in our
models can be fitted empirically in the lin-log diagram by the
function
\begin{equation} \log \dot{M}(t) = \log \dot{M}_0\,\frac{\mathrm{e}}{\tau}\,t\,\mathrm{e}^{-t/\tau}
\label{eq:fit}
\end{equation}
with time $t$ and the fit parameters $\log \dot{M}_0$ and $\tau$.
This holds for the following conditions: We shift the ordinate by
$\Delta \log \dot{M}/({\rm M_{\sun}\,yr^{-1}}) = +7$
and we consider accretion when $\log \dot{M} \ge -7$.
The zero point of the timescale is determined once the accretion rate exceeds $\log \dot{M} = -7$.
The fitted curves are plotted as thick lines in Fig.~\ref{accrcurve}.
Note that the ordinate displays the original values without the applied
shift.  If there are secondary accretion peaks, they are typically
ignored and levelled out by the routine. The overall quality of the
fit can be estimated by the standard deviation
\begin{equation}
\sigma = \sqrt{\frac{1}{n-1} \sum_{t=0}^{n} \left[ \dot{M}_\mathrm{fit}(t) - \dot{M}(t) \right]^2}
\label{eq:sigma}
\end{equation}
where $\dot{M}(t)$ is the actual value of $\dot{M}$ at the time $t$ from
our simulation, while $\dot{M}_\mathrm{fit}(t)$ denotes $\dot{M}$ calculated
using Eq.~(\ref{eq:fit}) for the same time.
The mean value of $\sigma$ for each model is given in Col.~9 of Table~\ref{tab:models}.
Prestellar cores where the fit routine fails or where $\sigma > 1$ are
not taken into account in our subsequent analysis.
This concerns a wide variety of cores, there is no correlation with the final mass
or the time of formation. However, they represent only
a small fraction of the total number of objects. The actual numbers of
fitted cores are listed in Col.~8 of Table~\ref{tab:models}.

\begin{figure*}
   \centering
   \includegraphics[width=17cm]{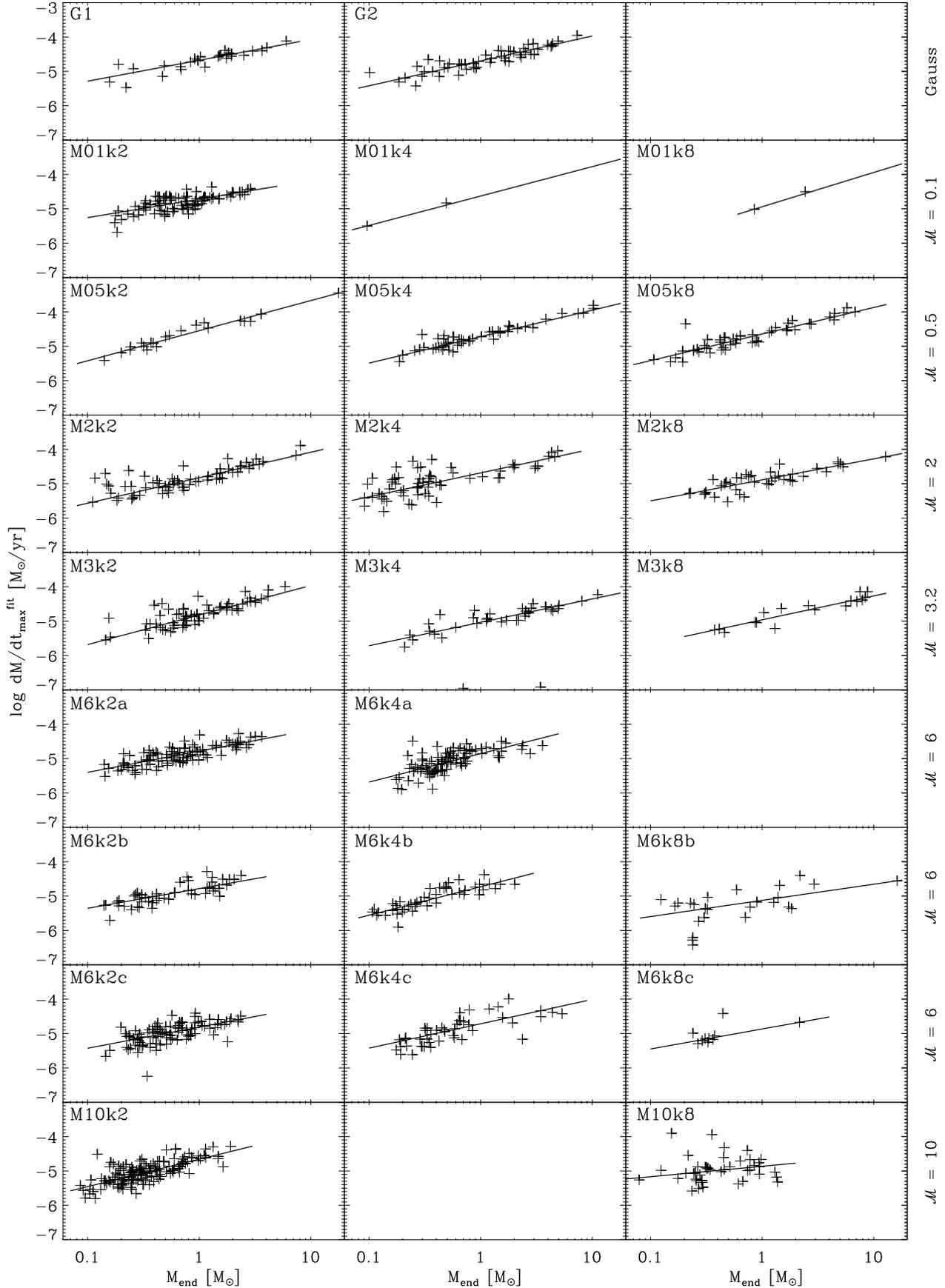}
   \caption{Peak accretion rates ($\dot{M}_\mathrm{max}^\mathrm{fit}$)
   versus $M_\mathrm{end}$ for all our models,
   sorted by Mach number \mach\ (top to bottom) and wave number $k$ (left to right).
   The straight line shows the applied linear fit.
   Details of the models can be found in Table~\ref{tab:models}.}
   \label{a0_a1}
   \end{figure*}

  When interpreting the fit parameter $\log \dot{M}_0$, the applied shift has
  to be taken into account.
   Thus, $\log \dot{M}_\mathrm{max}^\mathrm{fit} = \log (\dot{M}_0) - 7$ gives the
   real value of the peak accretion.
   This parameter is plotted for all
   protostellar cores and all models versus the respective final
   mass (Fig.~\ref{a0_a1}).  A correlation with $M_\mathrm{end}$ is
   obvious.  We apply a linear fit in the log-log diagram, which is
   indicated by the straight line.
   The fitted peak accretion rates show the same behaviour as the mean accretion rates \mmean.

\begin{figure*}
   \centering
   \includegraphics[width=17cm]{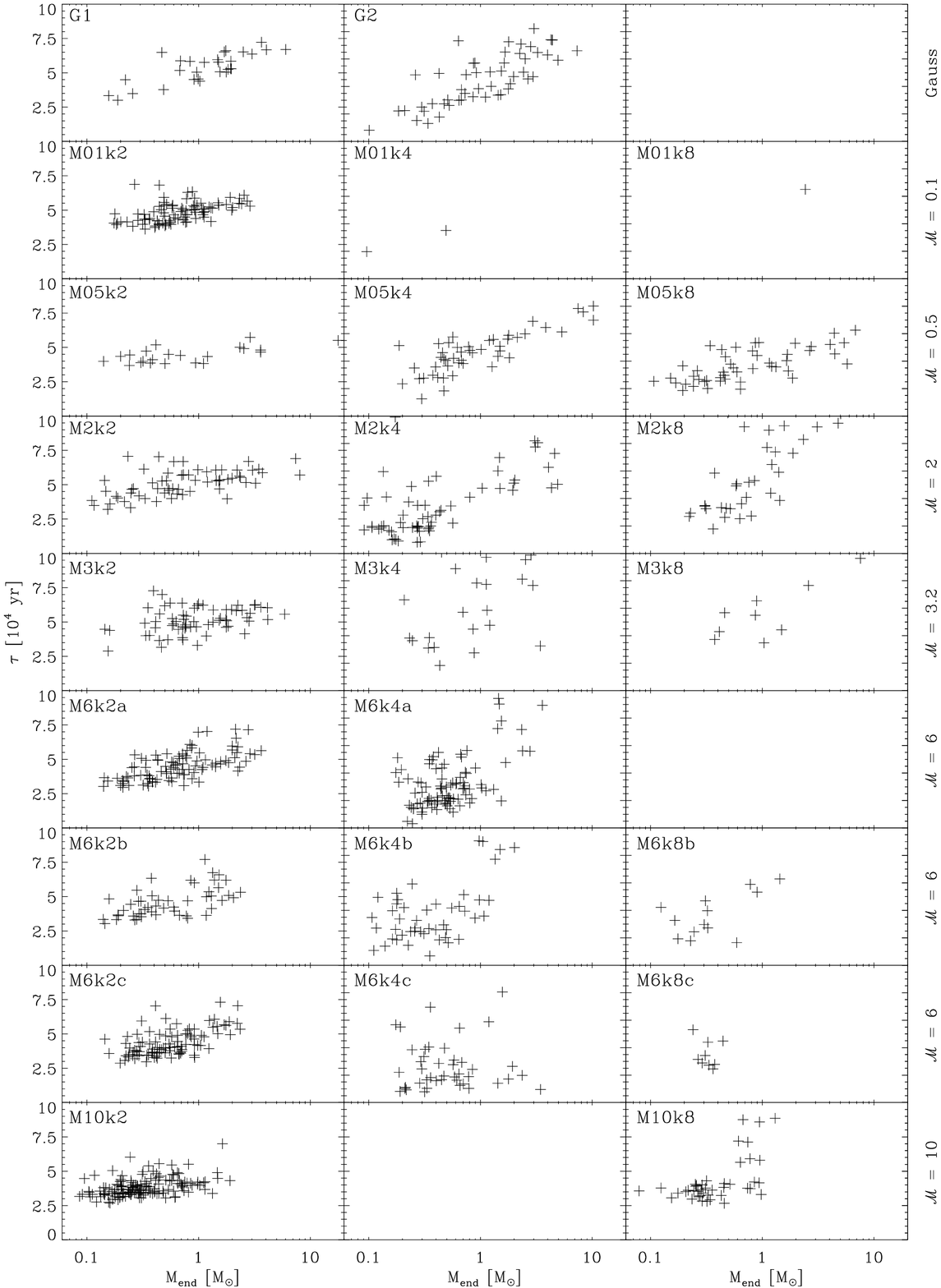}
   \caption{The time of maximum accretion $\tau$ for all models, arranged analogous to Fig.~\ref{a0_a1}.}
   \label{tau}
   \end{figure*}

   The parameter $\tau$ indicates the time of the maximum of the accretion curve.
   It is plotted for all protostellar cores in Fig.~\ref{tau}.
   In almost all models $\tau$ shows a correlation with the final mass.
   The parameter indicates how fast the gas falls in onto the core, therefore
   we expect it to be related to the local free-fall time and, thus,
   to the local density at the onset of collapse.
   It lies in the range $10^4 \lesssim \tau \lesssim 10^5$~yr, which is
   less than the global free-fall time $\bar{\tau}_\mathrm{ff}$.
   If we take an average value $\langle \tau \rangle \approx \bar{\tau}_\mathrm{ff}/3$,
   this suggests an initial overdensity of almost a factor of ten in the collapsing regions.


\subsection{Class~0 lifetimes and the effect of the turbulent medium}

   \begin{figure*}
   \centering
   \includegraphics[width=17cm]{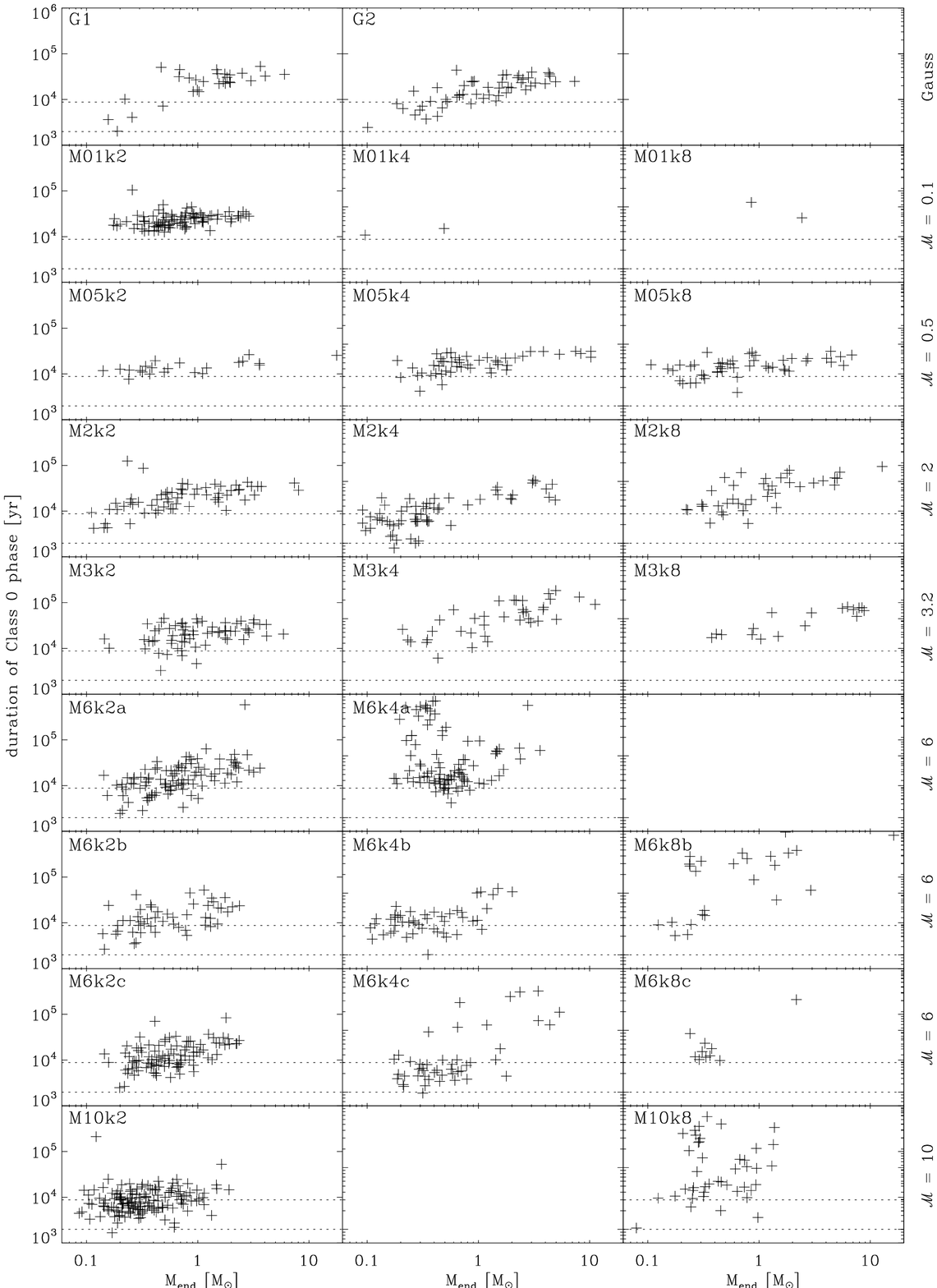}
   \caption{The assumed duration of Class 0 phase
   versus $M_\mathrm{end}$ for all models, arranged analogous to Fig.~\ref{a0_a1}.
   The dotted lines confine the range of this parameter according to observations
   (Andr\'{e} et al.\ \cite{andre00}).}
   \label{class01}
   \end{figure*}

   \begin{figure*}
   \centering
   \includegraphics[width=17cm]{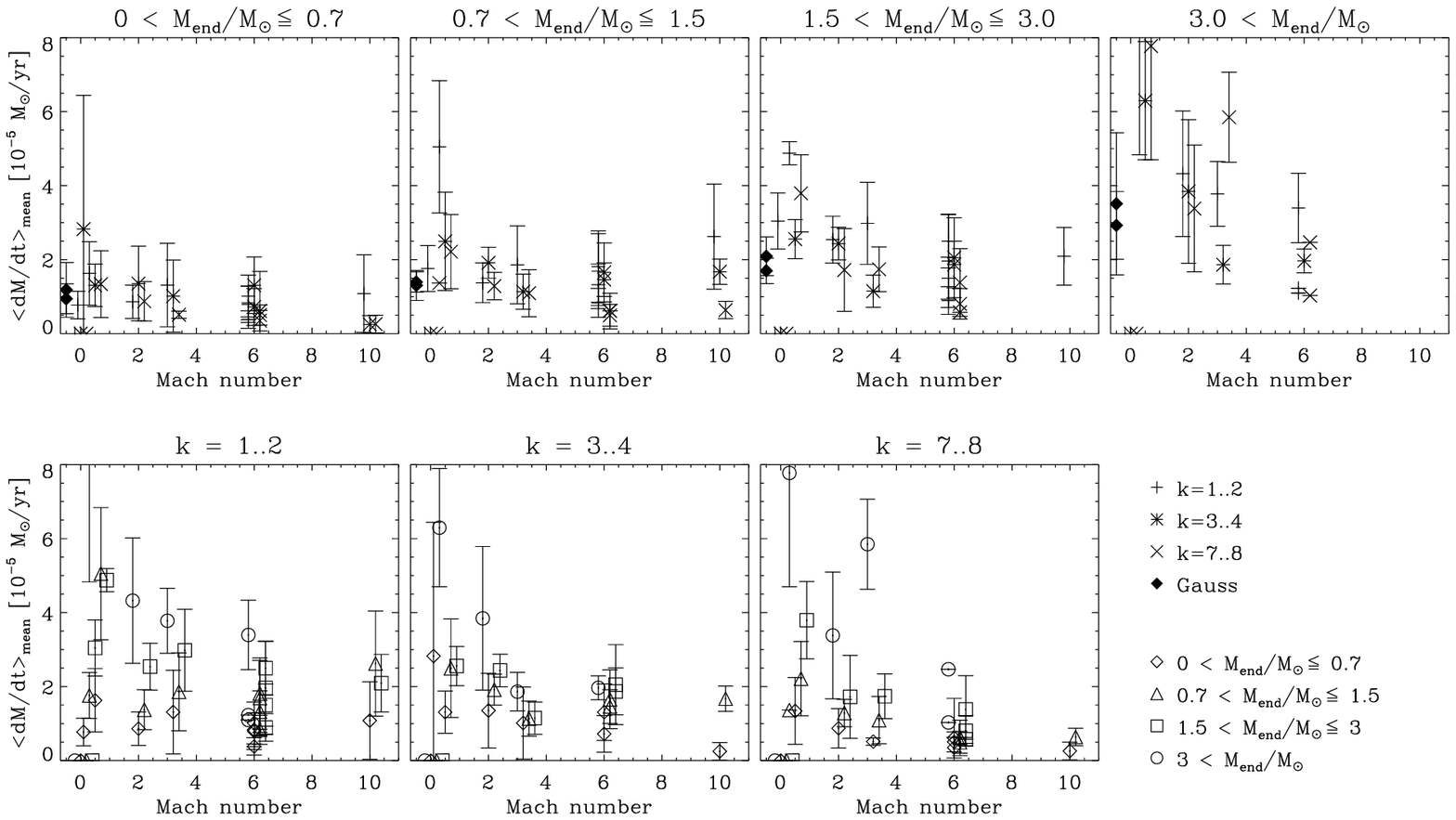}
   \caption{Averaged mean accretion rates \mmean$_\mathrm{mean}$ of all models
   versus Mach number.   For the sake of clarity the upper panel is split into
   mass bins, while the lower panel shows the same, separated according to
   the wave numbers.}
   \label{meanaccr_mach}
   \end{figure*}

We calculate the transition times from Class~0 to Class~I, assumed as described above.
This gives the duration of Class~0 phase for each protostar, the values are plotted
versus the corresponding final masses in Fig.~\ref{class01}.
The duration of Class~0 phase increases with increasing final mass.
Thus, a massive star is marked not only by a higher peak accretion rate but also
by a longer time spent in Class~0 phase.


The mean accretion rates \mmean\ of all individual protostellar cores of one model are
averaged in four mass bins: $0 < M_\mathrm{end}/{\rm M_{\sun}} \le 0.7$ (bin1), $0.7 <
M_\mathrm{end}/{\rm M_{\sun}} \le 1.5$ (bin2), $1.5 < M_\mathrm{end}/{\rm M_{\sun}} \le 3$ (bin3),
and $M_\mathrm{end}/{\rm M_{\sun}} > 3$ (bin4).
The values are given in Cols.~10 to 13 of Table~\ref{tab:models}.
Figure~\ref{meanaccr_mach} shows the relation of the averaged mean mass accretion rate
\mmean$_\mathrm{mean}$ to the Mach number for all models, split into mass bins and wave numbers, respectively.
Three conclusions can be drawn from the figure: Firstly, there is a trend that \mmean$_\mathrm{mean}$
decreases with increasing Mach number. That means that the mean accretion rate is lower, when
the rms velocity dispersion (i.e. the turbulent Mach number) of the medium is increased.
The stronger support of the turbulent medium against gravitational collapse typically results
in a lower mass accretion rate.
Secondly, \mmean$_\mathrm{mean}$ is higher for higher mass bins. This is consistent with the
findings for the individual mean and maximum accretion rates (\mmean\ and
$\dot{M}_\mathrm{max}^\mathrm{fit}$) discussed above.
Finally, though, there is no correlation of \mmean$_\mathrm{mean}$ with the wavenumber.
Apparently the scale of the driving energy has no influence on the accretion rate.


\section{\label{sect:comp_mod}Comparison with other theoretical models}

In the standard theory of isolated star formation (Shu \cite{shu77}),
which takes a singular isothermal sphere as initial condition,
the mass accretion rate is constant in time: $\dot{M} = 0.975\,c_\mathrm{s}^3/G$.
Note that also the Larson-Penston solution (Larson \cite{larson69}; Penston \cite{penston69}),
considering constant-density initial perturbations, gives a time-independent
accretion rate, however, at a higher level of $47\,c_\mathrm{s}^3/G$.
First numerical collapse calculations have been reported by Bodenheimer \& Sweigart
(\cite{bs68}), Larson (\cite{larson69}), and Hunter (\cite{hunter}).
Models with more realistic initial density profiles generally
predict accretion rates that decline with time:
The models of
Hunter (\cite{hunter}; \mmax\ $\approx 36\,c_\mathrm{s}^3/G$ from numerical
integrations of isothermal collapses),
Foster \& Chevalier (\cite{fost_chev}; \mmax\ $\approx 47\,c_\mathrm{s}^3/G$ from numerical
hydrodynamic simulations),
Tomisaka (\cite{tomisaka}; \mmax\ = $(4-40)\,c_\mathrm{s}^3/G$ from numerical MHD models),
Basu (\cite{basu}; \mmax\ = $13\,c_\mathrm{s}^3/G$ from a semi-analytical model),
Ogino et al.\ (\cite{ogino}; \mmax\ $\approx (30-230)\,c_\mathrm{s}^3/G$ from numerical hydrodynamic simulations),
Masunaga \& Inutsuka (\cite{masu_inu}; radiation hydrodynamic numerical codes),
Whitworth \& Ward-Thompson (\cite{wwt}), and
Motoyama \& Yoshida (\cite{moto_yosh}; \mmax\ $\gtrsim 42\,c_\mathrm{s}^3/G$)
predict mass accretion rates that peak shortly after the formation of the protostar
and decrease with time.
This shows better agreement with observational data
than constant accretion rates (see Sect.~\ref{sect:obs}).
Our results display the same behaviour and our values of \mmax\ $\approx (3-50)\,c_\mathrm{s}^3/G$
coincide quite well with those findings.
In contrast, some models yield mass accretion rates that increase with time
(McLaughlin \& Pudritz \cite {mcla_pud}; Bonnell et al.\ \cite{bonnell01a};
Behrend \& Maeder \cite{bm01}).

Theoretical models of star formation usually are scale-free. Thus, the results
strongly depend on the adopted physical scaling, e.g. the choice of the initial
density or temperature.
The comparison of numerical values of accretion rates therefore requires some care.
Again, in most cases the maximum accretion rates
scale approximately as a few times $M_\mathrm{J}/\tau_\mathrm{ff}$.

Whitworth \& Ward-Thompson (\cite{wwt}) presented an analytical model
for protostellar collapse using a Plummer-like density profile as
initial condition.  They successfully modelled the prestellar core
\object{L1544} in good agreement with observations.  Their \mmax
$\approx 8.1 \times 10^{-5}\ {\rm M_{\sun}\,yr^{-1}}$ corresponds
quite well to our Gaussian collapse cases for the same stellar mass (G1, G2).
However, the accretion history of the collapsing Plummer sphere cannot
be matched with our fit formula (\ref{eq:fit}). The increase is
steeper, while the decrease is slower compared to any of our models.
The slow decline might be due to the fact that Whitworth \& Ward-Thompson
(\cite{wwt}) use a not-truncated, infinite density profile, while our models have finite sizes.
A similar model was used by Motoyama \& Yoshida (\cite{moto_yosh}) who examined
the hypothesis that very high mass accretion rates exceeding $ 10^{-4}\
{\rm M_{\sun}\,yr^{-1}}$ require external triggering, as inferred from
some observations.  They find that the maximum accretion rate is
proportional to the momentum given to the cloud core in their
perturbed collapse model.  A momentum of $\gtrsim 0.1\ {\rm
  M_{\sun}\,km\,s^{-1}}$ causes an accretion rate of $\gtrsim 10^{-4}\
{\rm M_{\sun}\,yr^{-1}}$.

Smith (\cite{smith99}, \cite{smith00}) presented a formula for the mass
accretion rate with a sharp exponential rise and a power law decrease in time.
This model provides an early peak in which $\dot{M} \approx 10^{-4}\
{\rm M_{\sun}\,yr^{-1}}$ for $10^4$ years and eventually becoming $\dot{M} \approx 10^{-7}\
{\rm M_{\sun}\,yr^{-1}}$ for $10^6$ years.
However, his formula (Eq.~6 in Smith \cite{smith00}) applies to our models only
when choosing the parameters completely different than suggested, otherwise his accretion
curve has a more rapid increase but a slower decline than our models.

Bonnell et al.\ (\cite{bonnell01a}) analysed competitive accretion in
embedded stellar clusters by means of SPH simulations.  They find that
accretion in a cluster is highly non-uniform and that the accretion
rate is higher for stars near the cluster centre.  We do not see this
in our results, likely because the protostars in our model are at
different stages of evolution at a certain time, so this effect, if
existent, is superposed by the strongly time-dependent variation in
the accretion rate.  Also the evolution of $\dot{M}$ with time in
their models differs from our results: The mean accretion rate
reported by Bonnell et al.\ (\cite{bonnell01a}), determined from all
protostars in the cluster, increases with time
until near the end of the simulation when the gas is significantly
depleted.
The difference is probably caused by the different assumptions (e.g. lack
of turbulence, clustered potential). Indeed, the recent models of Bonnell \etal\
(\cite{bonnell03}) produce nearly constant accretion rates onto the
most massive stars in the cluster. The difference might be due to the fact
that the accretion rates are determined by the accretion onto the cluster
from outside, while in the models of Bonnell \etal\ (\cite{bonnell01a})
all the mass was already in the cluster.

Reid et al.\ (\cite{reid}) used a logatropic equation of state as
basis for their hydrodynamical simulations of isolated star formation.
Their accretion rate depends on the size of the core. It increases cubically and reaches
the maximum when the expansion wave leaves the core, then it falls steadily.
With the adopted scaling, \mmax\ is one to two orders of magnitude smaller
than in our models. Consequently the whole accretion process lasts much longer,
several $10^6$ years, which
is in contradiction to estimates of rapid star formation
(Elmegreen \cite{elmeg00}; Hartmann \etal\ \cite{hbb01}; Hartmann
\cite{hartmann03}; Mac Low \& Klessen \cite{mor_ralf}).

Wuchterl \& Tscharnuter (\cite{wuch_tscha}) find from models based on
radiation hydrodynamics time varying accretion rates of a few
$10^{-6}\ {\rm M_{\sun}\,yr^{-1}}$ for the phase $<
0.8\,M_\mathrm{end}$, which is about an order of magnitude lower than
our values, especially for the Gaussian collapse or large-scale turbulence.
The reason may be that protostellar cores in our models form by external
compression before gravity takes over. This results in enhanced accretion
rates relative to cores that begin contraction close to hydrostatic equilibrium.

Hennebelle et al.\ (\cite{henne}) performed numerical simulations
where the collapse is triggered by a steady increase in the external
pressure.  \mmax\ is reached immediately after the formation of the
central protostar (i.e. during Class~0 phase), followed by a steady
decrease to the Class~I phase. The more rapid and the more prolonged
the increase in external pressure, the higher is \mmax, ranging from
$6.5 \times 10^{-6}\ {\rm M_{\sun}\,yr^{-1}}$ to $2.6 \times 10^{-5}\
{\rm M_{\sun}\,yr^{-1}}$, corresponding to $\sim (4-16)\,c_\mathrm{s}^3/G$. 
The qualitative behaviour of the accretion process
does not differ much from our models, but the peak accretion rates are slightly
smaller except in their models with the most rapid compression.

\section{\label{sect:obs}Comparison with observations}

It is very difficult to measure mass accretion rates directly from observations
(e.g. from inverse P~Cygni profiles), instead they often have to be inferred indirectly
based on the spectral energy distributions
(SEDs) of protostars or using outflow characteristics (Hartigan \etal\ \cite{hartigan95};
Bontemps \etal\ \cite{bontemps}).  The
 correlation between accretion rates and outflow strength, however, is
 still subject of strong debate (Wolf-Chase et al.\ \cite{wolf}).
For Class 0 objects typical
 mass accretion rates are estimated in the range $10^{-5} \lesssim \dot{M}/{\rm
 M_{\sun}\,yr^{-1}} \lesssim 10^{-4}$ (Hartmann \cite{hartmann98}; Narayanan
\etal\ \cite{naray98}; Andr\'{e}
et al.\ \cite{andre99}; Ceccarelli et al.\ \cite{cecc}; Jayawardhana
et al.\ \cite{jaya}; Di Francesco et al.\ \cite{difranc}; Maret et
al.\ \cite{maret}; Beuther \etal\ \cite{beuther02a}, \cite{beuther02b}).
The growth rate of Class~I objects is
believed to be about an order of magnitude smaller
(Henriksen et al.\ \cite{henriksen}; Andr\'{e} et al.\
\cite{andre00}), with observational values
between $\sim 10^{-7}$ and $\sim 5 \times 10^{-6}\
\mathrm{M_{\sun}\,yr^{-1}}$ (Brown \& Chandler \cite{brown_chand};
Greene \& Lada \cite{greene_lada}; Boogert \etal\ \cite{bhb02};
Yokogawa et al.\ \cite{yoko}; Young \etal\ \cite{young03}).

Bontemps et al.\ (\cite{bontemps}) studied the outflow activities in a
sample of 45 low-mass embedded young stellar objects. They estimate
that the observed decline of CO outflow momentum fluxes with time
results from a decrease of the mass accretion rate from $\sim 10^{-5}\
{\rm M_{\sun}\,yr^{-1}}$ for the youngest Class 0 protostars to $\sim
10^{-7}\ {\rm M_{\sun}\,yr^{-1}}$ for the most evolved Class I
objects.
Furthermore, they propose a simple exponential dependency of the
accretion rate with time:
$\dot{M} = (M_\mathrm{env}^0 / \tau) \mathrm{e}^{-t/\tau}$
with initial mass of the dense clump $M_\mathrm{env}^0$ and a
characteristic time $\tau \approx 9 \times 10^4\,\mathrm{yr}$.
This is comparable to our Eq.~(\ref{eq:fit}).
A similar exponential equation is also used by Myers \etal\ (\cite{myers98}),
while Henriksen et al.\ (\cite{henriksen}) describe the accretion
rate by an equation that asymptotically approaches a power-law
dependence at late times.

Brown \& Chandler (\cite{brown_chand}), who determined an
upper limit of $\dot{M}\lesssim (2-4) \times 10^{-7}\ {\rm
  M_{\sun}\,yr^{-1}}$ for two Class~I protostars in Taurus, also
conclude that the accretion rate is not constant in time and likely
much higher in the early phase.
On the other hand, Hirano et al.\ (\cite{hirano}) observed a dozen of
deeply embedded young stellar objects of both Class~0 and I type
and derived the same mass accretion rates of $(1-5) \times 10^{-6}\ {\rm M_{\sun}\,yr^{-1}}$
for all of them. Unlike other authors, they argue that there is no
significant difference in $\dot{M}$ between Class~0 and Class~I sources.

The values given above correspond to the accretion rates
derived for the model of gravoturbulent star formation discussed here.
They also decrease from $10^{-5}$ to $10^{-4}\ {\rm M_{\sun}\,yr^{-1}}$
during Class~0 phase to less than $10^{-7}\ {\rm M_{\sun}\,yr^{-1}}$ in later
stages.  However, the supposed transition between Class~0 and Class~I
takes place still during the peak accretion phase.  The accretion
rates in our models typically do not decline significantly until about 80\% of
the final mass have been accreted (Fig.~\ref{accrcurve}).  This is
unlike e.g. the model of Reid et al.\ (\cite{reid}), where $\dot{M}$
begins to fall off when about half of the mass of the core has been
accreted.
Given the uncertainties of the mass estimate for Class 0/I transition
we do not consider this a large discrepancy.

According to observations, Class~0 objects have an estimated
 lifetime of $\sim (1 - 3) \times 10^4$~yr
(Andr\'{e} et al.\ \cite{andre00}).  In our models this parameter is
widely spread, ranging from $\sim 10^4$ to $> 10^5$\,yr,
but for a 1~M$_{\sun}$ star it lies roughly in the range
deduced from observations (see Fig.~\ref{class01}).


\section{Summary}
\label{sec:summary}

We have studied protostellar mass accretion rates from numerical models
of star formation based on gravoturbulent fragmentation.
Twenty-four models covering a wide range of environmental conditions from low to
high turbulent velocities and different driving scales with a total number of
1325 protostellar cores have been investigated.
Our main results may be summarised as follows:

\begin{enumerate}
\item
An order-of-magnitude estimate for mass accretion rates resulting from gravoturbulent
fragmentation is given by $\dot{M} \approx M_\mathrm{J} / \tau_\mathrm{ff}$ with $M_\mathrm{J}$
being the mean thermal Jeans mass and $\tau_\mathrm{ff}$ the corresponding free-fall time.
\item
However,
protostellar mass accretion is a highly time-variant process. It can
be approximated by the empirical function
$\log \dot{M}(t) = \log \dot{M}_0\,(\mathrm{e}/\tau)\,t\,\mathrm{e}^{-t/\tau}$.
The peak accretion rate is reached during Class 0 stage, shortly after the formation of 
the core; its value ranges between about $5 \times 10^{-6}$ and $10^{-4}\ {\rm M_{\sun}\,yr^{-1}}$.
The maximum accretion rate is approximately one order of magnitude higher than the constant rate
predicted by the collapse of a classical singular isothermal sphere.

\item Around the peak accretion phase the mass accretion rates are roughly constant.
The mean accretion rates 
increase with increasing final mass.
More massive stars have higher mass accretion rates and tend to form first.

\item The same applies to the fitted peak accretion rates, which are also proportional to the final stellar mass.

\item There is a similar correlation between the duration of Class~0 phase (assuming that 
half of the final mass is accreted in this phase) and the final mass.
\item \mmean$_\mathrm{mean}$ decreases with increasing Mach number of the turbulent
environment, but is not correlated with the driving wavenumber.
\end{enumerate}

\noindent Our results agree well with many other models concerning the time evolution
of the mass accretion process and the value of the peak accretion rate.
In particular, the accretion rates from our models show an exponential decline, as it 
is also proposed by Bontemps et al.\ (\cite{bontemps}), Myers et al.\ (\cite{myers98}) 
and Smith (\cite{smith99}, \cite{smith00}).
They also match observational findings like the supposed decline of the mass accretion rate
from Class~0 to Class~I phase.
We conclude that a theory of star formation based on gravoturbulent fragmentation of
molecular clouds is an adequate approach to describe stellar birth in the Milky Way.

\begin{acknowledgements}
This work was supported by the Emmy Noether Programme of the {\em Deutsche
Forschungsgemeinschaft} (grant no.\ KL1358/1).
We are grateful to Dirk Froebrich and Philippe Andr\'{e} for helpful remarks.
We also wish to thank the referee Ian Bonnell for his
insightful comments and suggestions.
\end{acknowledgements}


\begin{thebibliography}{}

\bibitem[2001]{adams_myers} Adams, F.~C., \& Myers, P.~C. 2001, \apj, 553, 744

\bibitem[1999]{andre99} Andr\'{e}, P., Motte, F., \& Bacmann, A. 1999, \apj, 513, L57

\bibitem[2000]{andre00} Andr\'{e}, P., Ward-Thompson, D., \& Barsony, M. 2000, in 
Proto\-stars and Planets IV, ed.\ V. Mannings, A.~P. Boss, \& S.~S. Russell 
(Tucson: University of Arizona Press), 59

\bibitem[2002]{bpml02} Ballesteros-Paredes, J., \& Mac~Low, M.-M. 2002, \apj, 570, 734

\bibitem[1999a]{bvs99} Ballesteros-Paredes, J., V\'azquez-Semadeni, E., \& Scalo,
J. 1999a, \apj, 515, 286

\bibitem[1999b]{bhv99} Ballesteros-Paredes, J., Hartmann, L., \&
V\'azquez-Semadeni, E. 1999b, \apj, 527, 285

\bibitem[1997]{basu} Basu, S. 1997, \apj, 485, 240

\bibitem[1997]{bb97} Bate, M.~R., \& Burkert, A. 1997, \mnras, 288, 1060

\bibitem[1995]{bbp95} Bate, M.~R., Bonnell, I.~A., \& Price, N.~M. 1995, \mnras, 277, 362

\bibitem[2002]{bbb02} Bate, M.~R., Bonnell, I.~A., \& Bromm, V. 2002, \mnras, 332, L65

\bibitem[2001]{bm01} Behrend, R., \& Maeder, A. 2001, \aap, 373, 190

\bibitem[1990]{benz90} Benz, W. 1990, in The Numerical Modelling of Nonlinear Stellar Pulsations,
ed.\ J. R. Buchler (Dordrecht: Kluwer), 269

\bibitem[2002a]{beuther02a} Beuther, H., Schilke, P., Sridharan, T.~K., \etal\
2002a, \aap, 383, 892

\bibitem[2002b]{beuther02b} Beuther, H., Schilke, P., Gueth, F., et al. 2002b, \aap, 387, 931

\bibitem[1968]{bs68} Bodenheimer, P., \& Sweigart, A. 1968, \apj, 152, 515

\bibitem[1997]{bonnell97} Bonnell, I.~A., Bate, M.~R., Clarke, C.~J., \& Pringle, J.~E. 1997, \mnras, 285, 201

\bibitem[2001a]{bonnell01a} Bonnell, I.~A., Bate, M.~R., Clarke, C.~J., \& Pringle, J.~E. 2001a, \mnras, 323, 785

\bibitem[2001b]{bonnell01b} Bonnell, I.~A., Clarke, C.~J., Bate, M.~R., \& Pringle, J.~E. 2001b, \mnras, 324, 573

\bibitem[2003]{bonnell03} Bonnell, I.~A., Bate, M.~R., \& Vine, S.~G. 2003, \mnras, 343, 413

\bibitem[1996]{bontemps} Bontemps, S., Andr\'{e}, P., Terebey, S., \& Cabrit, S. 1996, \aap, 311, 858

\bibitem[2002]{bhb02} Boogert, A.~C.~A., Hogerheijde, M.~R., \& Blake, G.~A. 2002, \apj, 568, 761

\bibitem[2001]{bourke} Bourke, T.~L., Myers, P.~C., Robinson, G., \& Hyland, A.~R. 2001, \apj, 554, 916

\bibitem[1999]{brown_chand} Brown, D.~W., \& Chandler, C.~J. 1999, \mnras, 303, 855

\bibitem[2000]{cecc} Ceccarelli, C., Castets, A., Caux, E., et al.\ 2000, \aap, 355, 1129

\bibitem[1999]{clarke99} Clarke, C.~J. 1999, \mnras, 307, 328

\bibitem[1999]{crutcher} Crutcher, R.~M. 1999, \apj, 520, 706

\bibitem[2001]{difranc} Di Francesco, J., Myers, P.~C., Wilner, D.~J., Ohashi, N., \& Mardones, D. 2001, \apj, 562, 770

\bibitem[1993]{elmeg93} Elmegreen, B.~G. 1993, \apj, 419, L29

\bibitem[2000]{elmeg00} Elmegreen, B.~G. 2000, \apj, 530, 277

\bibitem[2002]{elmeg02} Elmegreen, B.~G. 2002, \apj, 577, 206

\bibitem[1993]{fost_chev} Foster, P.~N., \& Chevalier, R.~A. 1993, \apj, 416, 303

\bibitem[2002]{greene_lada} Greene, T.~P., \& Lada, C.~J. 2002, \aj, 124, 2185

\bibitem[1995]{hartigan95} Hartigan, P., Edwards, S., \& Ghandour, L. 1995, \apj, 452, 736

\bibitem[1998]{hartmann98} Hartmann, L. 1998, Accretion processes in star
formation (Cam\-bridge: Cambridge University Press)

\bibitem[2003]{hartmann03} Hartmann, L. 2003, \apj, 585, 398

\bibitem[2001]{hbb01} Hartmann, L., Ballesteros-Paredes, J., \& Bergin, E.~A. 2001, \apj, 562, 852

\bibitem[2001a]{hmk01} Heitsch, F., Mac~Low, M.-M., \& Klessen, R.~S. 2001a, \apj, 547, 280

\bibitem[2001b]{heitsch01} Heitsch, F., Zweibel, E.~G., Mac~Low, M.-M., Li, P., \& Norman, M.~L. 2001b, \apj, 561, 800

\bibitem[2003]{henne} Hennebelle, P., Whitworth, A.~P., Gladwin, P.~P., \& Andr\'{e}, P. 2003, \mnras, 340, 870

\bibitem[1997]{henriksen} Henriksen, R., Andr\'e, P., \& Bontemps, S. 1997, \aap, 323, 549

\bibitem[2003]{hirano} Hirano, N., Ohashi, N., Dobashi, K., Shinnaga, H., \& Hayashi, M. 2003,
in Proc.\ of the IAU 8th Asian-Pacific Regional Meeting, Vol.~II, eds.\ S. Ikeuchi, J. Hearnshaw,
\& T. Hanawa, ASP Conf.\ Ser., 289, 141

\bibitem[1971]{hws71} Hollenbach, D.~J., Werner, M.~W., \& Salpeter, E.~E. 1971, \apj, 163, 165

\bibitem[1977]{hunter} Hunter, C. 1977, \apj, 218, 834

\bibitem[1982]{hunter_fleck} Hunter, J.~H., Jr., \& Fleck, R.~C., Jr. 1982, \apj, 256, 505

\bibitem[2001]{jaya} Jayawardhana, R., Hartmann, L., \& Calvet, N. 2001, \apj, 548, 310

\bibitem[1997]{klessen97} Klessen, R.~S. 1997, \mnras, 292, 11

\bibitem[2001a]{klessen01a} Klessen, R.~S. 2001a, \apj, 550, L77

\bibitem[2001b]{klessen01b} Klessen, R.~S.  2001b, \apj, 556, 837

\bibitem[2000]{kless_bur00} Klessen, R.~S., \& Burkert, A. 2000, \apjs, 128, 287

\bibitem[2001]{kless_bur01} Klessen, R.~S., \& Burkert, A. 2001, \apj, 549, 386

\bibitem[2000]{khm00} Klessen, R.~S., Heitsch, F., \& Mac Low, M.-M. 2000, \apj, 535, 887

\bibitem[2003]{ladalada} Lada, C.~J., \& Lada, E.~A. 2003, \araa, 41, 57

\bibitem[2000]{langer00} Langer, W.~D., van Dishoeck, E.~F., Bergin, E.~A., \etal\ 2000, in
Proto\-stars and Planets IV, ed.\ V. Mannings, A.~P. Boss, \& S.~S. Russell
(Tucson: University of Arizona Press), 29

\bibitem[1969]{larson69} Larson, R.~B. 1969, \mnras, 145, 271

\bibitem[1985]{larson85} Larson, R.~B. 1985, \mnras, 214, 379

\bibitem[2003]{larson03} Larson, R.~B. 2003, Rep.\ Prog.\ Phys., 66, 1651

\bibitem[1999]{lee_my} Lee, C.~W., \& Myers, P.~C. 1999, \apjs, 123, 233

\bibitem[1993]{lp93} Lubow, S.~H., \& Pringle, J.~E. 1993, \mnras, 263, 701

\bibitem[1999]{maclow99} Mac~Low, M.-M. 1999, \apj, 524, 169

\bibitem[2002]{maclow02} Mac~Low, M.-M. 2002, \apss, 281, 429

\bibitem[2004]{mor_ralf} Mac Low, M.-M., \& Klessen, R.~S. 2004, Rev.\ Mod.\ Phys., 76, 125

\bibitem[2000]{maclow_oss00} Mac~Low, M.-M., \& Ossenkopf, V. 2000, \aap, 353, 339

\bibitem[1998]{maclow98} Mac~Low, M.-M., Klessen, R.~S., Burkert, A., \& Smith, M.~D. 1998,
Phys.\ Rev.\ Lett., 80, 2754

\bibitem[2003]{mak03} Mac~Low, M.-M., de Avillez, M.~A., \& Korpi,
  M.~J. 2003, in How Does the Galaxy Work?, eds. E.~J.~Alfaro,
  E.~P{\'e}rez, \& J.~Franco (Dordrecht: Kluwer), in press (astro-ph/0310817)

\bibitem[2002]{maret} Maret, S., Ceccarelli, C., Caux, E., Tielens, A.~G.~G.~M., \& Castets, A. 2002, \aap, 395, 573

\bibitem[2000]{masu_inu} Masunaga, H., \& Inutsuka, S. 2000, \apj, 531, 350

\bibitem[1997]{mcla_pud} McLaughlin, D.~E., \& Pudritz, R.~E. 1997, \apj, 476, 750

\bibitem[1992]{monagh92} Monaghan, J.~J. 1992, \araa, 30, 543

\bibitem[2003]{moto_yosh} Motoyama, K., \& Yoshida, T. 2003, \mnras, 344, 461

\bibitem[1998]{man98} Motte, F., Andr{\'e}, P., \& Neri, R. 1998, \aap, 336, 150

\bibitem[1998]{myers98} Myers, P.~C., Adams, F.~C., Chen, H., \& Schaff, E. 1998, \apj, 492, 703

\bibitem[1998]{nakano98} Nakano, T. 1998, \apj, 494, 587

\bibitem[1998]{naray98} Narayanan, G., Walker, C.~K., \& Buckley, H.~D. 1998, \apj, 496, 292

\bibitem[1999]{ogino} Ogino, S., Tomisaka, K., \& Nakamura, F. 1999, \pasj, 51, 637

\bibitem[2002]{oss_maclow02} Ossenkopf, V., \& Mac~Low, M.-M. 2002, \aap, 390, 307

\bibitem[1995]{padoan95} Padoan, P. 1995, \mnras, 277, 377

\bibitem[1999]{pado_nord99} Padoan, P., \& Nordlund, \AA. 1999, \apj, 526, 279

\bibitem[2002]{pado_nord02} Padoan, P., \& Nordlund, \AA. 2002, \apj, 576, 870

\bibitem[1969]{penston69} Penston, M.~V. 1969, \mnras, 144, 425

\bibitem[2002]{reid} Reid, M.~A., Pudritz, R.~E., \& Wadsley, J. 2002, \apj, 570, 231

\bibitem[1977]{shu77} Shu, F.~H. 1977, \apj, 214, 488

\bibitem[1987]{sal87} Shu, F.~H., Adams, F.~C., \& Lizano, S. 1987, \araa, 25, 23

\bibitem[1999]{smith99} Smith, M.~D. 1999, \apss, 261, 169

\bibitem[2000]{smith00} Smith, M.~D. 2000, Ir.\ Astron.\ J., 27, 25

\bibitem[1998]{sog98} Stone, J.~M., Ostriker, E.~C., \& Gammie, C.~F. 1998, \apj, 508, L99

\bibitem[1996]{tmw96} Taylor, S.~D., Morata, O., \& Williams, D.~A. 1996, \aap, 313, 269

\bibitem[1996]{tomisaka} Tomisaka, K. 1996, \pasj, 48, L97

\bibitem[1998]{dis_blake} van Dishoeck, E.~F., \& Blake, G.~A. 1998, \araa, 36, 317

\bibitem[2004]{vsb03} V\'azquez-Semadeni, E., Shadmehri, M., \& Ballesteros-Paredes, J.
2004, \apj, submitted (astro-ph/0208245)

\bibitem[2002]{vrc02} Visser, A.~E., Richer, J.~S., \& Chandler, C.~J. 2002, \aj, 124, 2756

\bibitem[2001]{wwt} Whitworth, A.~P., \& Ward-Thompson, D. 2001, \apj, 547, 317

\bibitem[2003]{wolf} Wolf-Chase, G., Moriarty-Schieven, G., Fich, M., \& Barsony, M. 2003, \mnras, 344, 809

\bibitem[2001]{wucht_kless01} Wuchterl, G., \& Klessen, R.~S. 2001, \apj, 560, L185

\bibitem[2003]{wuch_tscha} Wuchterl, G., \& Tscharnuter, W.~M. 2003, \aap, 398, 1081

\bibitem[2003]{yoko} Yokogawa, S., Kitamura, Y., Momose, M., \& Kawabe, R. 2003, in
Proc.\ of the IAU 8th Asian-Pacific Regional Meeting, Vol.~II, eds.\ S. Ikeuchi, J. Hearnshaw, \& T. Hanawa,
ASP Conf.\ Ser., 289, 239

\bibitem[2003]{young03} Young, C.~H., Shirley, Y.~L., Evans, N.~J., II, \& Rawlings, J.~M.~C. 2003, \apjs, 145, 111

\end{thebibliography}
\end{document}